\documentstyle[12pt]{article}
\begin{document}

\input epsf

\begin{center}
   {\Large \bf Strangeness and Charm Signatures \\
of the Quark-Gluon Plasma}
   \end{center}

\vspace{0.2cm}

\begin{center}

{\bf Mark I. Gorenstein}

\vspace{0.2cm}
Bogolyubov Institute for 
Theoretical Physics,
Kiev, Ukraine\\ and\\
Institut f\"ur Theoretische Physik, Universit\"at  Frankfurt,
Germany 

\end{center}

\vspace{0.5cm}
\noindent
{\bf Abstract.} Strangeness,
charmonium and open charm yields in relativistic nucleus-nucleus
collisions are considered within statistical model approach
as potential signals of the quark-gluon plasma.   

\vspace{0.5cm}
\noindent
{\bf I. Introduction}

The present status of the quark-qluon plasma (QGP) 
in nucleus--nucleus (A+A) 
collisions is somewhat uncertain.  
Experimental data for A+A collisions with truly heavy beams have become 
available: Au+Au at 11~A$\cdot$GeV at the BNL AGS and Pb+Pb at 
158~A$\cdot$GeV 
at the CERN SPS \cite{QM96}.
A systematic analysis of these data could yield clues to whether
a short-lived phase with quark and gluon constituents, the QGP,
exists during the hot and dense stage of these reactions.
This question, whether the QGP is already produced with the 
currently operating heavy ion accelerators, is right now vigorously
debated.
There are also convincing hopes to find new reliable evidences of the 
QGP within a few years. The accelerators of a new generation
RHIC BNL and LHC CERN will start soon to operate.

Since a long time \cite{Fe:50} statistical models
are used to describe hadron multiplicities
in high energy collisions.
Thermal hadron production models have been successfully
used to fit the data on particle multiplicities
in A+A collisions at the CERN SPS energies (see, e.g.
\cite{Go:97,Be:98,Go:99}).
Due to the large
number of particles a grand canonical formulation
is used for the modeling of high energy heavy
ion collisions. 
Recently, an impressive success of the statistical model
applied to hadron multiplicities in elementary
$e^+ + e^-$, $p+p$ and $p+\bar{p}$ interactions at
high energy was also reported \cite{Be:96}.
However,
in the latter case the  use of a canonical formulation of the
model, which assures the exact conservation of the
conserved charges, is necessary (see, e.g., \cite{Cl} and references 
therein). 

The temperature parameter which characterizes
the available phase space
for the hadron production
is found in these interactions to be 160--190 MeV \cite{Be:96}.
It does not show any significant dependence on the type
of reaction and on the collision energy (at the SPS energies and higher).
This fact suggests the  possibility to 
ascribe the observed statistical properties
of hadron production systematics in elementary and nuclear collisions
at high energies to  the statistical nature
of the hadronization process
\cite{Be:96,Be:98,St:99}.

\vspace{0.5cm}
\noindent
{\bf II. Strangeness Production}

The enhanced production of
strangeness was considered by many authors as a potential
signal of QGP formation (see, e.g., Ref.~\cite{koch86}).
The expectation was that strangeness production should
rapidly increase when the energy transition region is crossed from below.
The strangeness to pion ratio is indeed observed to increase
in A+A collisions. It seems however that this is not a signal of the QGP.
The low level of strangeness production in p+p interactions
as compared to the strangeness yield in central A+A collisions,
called strangeness enhancement,
can be also understood to a large extent as due to the effect of 
the exact strangeness 
conservation. The canonical ensemble treatment
of the strangeness conservation leads to the
additional suppression factors imposed on the strange hadrons
production in small systems created in p+p collisions \cite{Be:96,Cl}. 
Another important point is that for the chemical freeze-out parameters,  
temperature $T$ and baryonic chemical potential $\mu_b$,
found for the SPS energies the strangeness to entropy ratio  
is larger in the equilibrium HG than in the equilibrium QGP 
\cite{Go:99}.

To estimate the strangeness to entropy
ratio let us consider the quantity
\begin{equation}\label{sentr}
R_s~\equiv~\frac{N_s+N_{\bar{s}}}{S}~,
\end{equation}
where $N_s$ and $N_{\bar{s}}$ are the numbers of strange quarks and 
antiquarks, and S is the total entropy of the system. In the QGP we use the 
ideal gas approximation of 
massless {\it u}-, {\it d}-(anti)quarks and gluons, strange 
(anti)quarks  with $m_s\cong 150$ MeV and (anti)charm quarks with 
$m_c\cong 1500$~MeV. For the HG state the values of $N_s$ and $N_{\bar{s}}$
are calculated as a sum of all $s$ and $\bar{s}$ inside hadrons,
and $S$ is the total HG entropy.
The behaviour of $R_s$ (\ref{sentr}) for the HG and QGP
is shown in Fig.~1 as a function of $T$ for $\mu_B=0$.
In the wide range of $T=200\div 500$~MeV
one finds an almost
constant value of $R_s$ in the QGP which is smaller than the corresponding
quantity in the HG.
The total entropy as well as
the total number of strange quarks and antiquarks are expected to 
be conserved approximately during the hadronization of QGP.  This
suggests that the value of $R_s$ at the HG chemical freeze-out
should be close to that in the equilibrium QGP and smaller than in the 
HG at chemical equilibrium.  
Therefore, the strangeness {\it suppression} in the
HG would become a signal for the formation of QGP at the early stage of
$A$+$A$ collision at the CERN SPS energies.  

The same conclusion was
obtained in Ref.~\cite{Ga1:98}.
In the model presented in that paper it was assumed that
due to the statistical nature
of the  creation process
the strangeness in the
early stage is already in equilibrium and therefore possible secondary
processes do not modify its value.
As the strangeness to entropy ratio is lower
in the QGP than in
the confined matter,
the {\it suppression} of strangeness production is expected to
occur when crossing the transition energy range from below.
The total strangeness production is usually studied using
the experimental ratio
\begin{equation}\label{Es}
E_s~=~\frac{\langle \Lambda \rangle + \langle K + 
\bar{K}\rangle}{\langle \pi \rangle }~, 
\end{equation}
which measures the ratio of the mean multiplicities 
of $\Lambda$ hyperons and $K,\bar{K}$ mesons to the multiplicity of
pions. This is an
experimental analog of the quantity $R_s$ (\ref{sentr}).
In Fig.~2 the experimental data of $E_s$ ratio (\ref{Es})
are shown together with its theoretically expected behaviour
according to Ref.~\cite{Ga1:98}. The deconfinement phase transition
causes a nonmonotonic
behaviour of $E_s$ with the collision energy.
The maximum of $E_s$
is in the energy region between the AGS and SPS.

\vspace{0.5cm}
\noindent
{\bf III. Charmonium and Open Charm Production}

Charmonium production in 
hadronic \cite{Ma:95} and nuclear \cite{Vo:99} collisions
is usually considered to be  composed of three  stages:
the creation of a $c\overline{c}$ pair, the formation of a
bound $c\overline{c}$ state and the subsequent interaction 
of this  $c\overline{c}$ bound state with the
surrounding matter. 
The first process is calculated within perturbative QCD,
whereas modeling of  non--perturbative dynamics is needed to describe 
the last two stages (see, e.g., \cite{Ge:98} and references therein).
The interaction of the bound $c\overline{c}$ state with  matter
causes suppression of the finally observed charmonium yield
relative to the initially created number of bound $c\overline{c}$
states. 
This initial number is assumed to be proportional to the number
of Drell--Yan lepton pairs, which then allows for the experimental study
of the charmonium suppression pattern.
It was proposed
\cite{Sa:86} that the magnitude of the measured suppression
in nuclear collisions can be used as a probe of the state of
high density matter created at the early stage of the collision.
The rapid increase of the suppression
({\it anomalous suppression}) observed when going from peripheral
to central Pb+Pb collisions \cite{NA50} 
is often attributed to the formation
of the QGP
\cite{Sa:97}. 

It was recently found \cite{Ga1:98,Ga2:98} 
that the mean 
multiplicity of $J/\psi$ mesons increases proportionally to the mean
multiplicity of pions when p+p, 
p+A
and A+A collisions at CERN SPS energies are 
considered.
We illustrate this unexpected experimental fact by reproducing in
Fig.~3  
the ratio $\langle J/\psi \rangle / \langle h^- \rangle$ 
as a function of the mean number of nucleons participating
in the interaction for inelastic nuclear collisions at the
CERN SPS. 
The $\langle J/\psi \rangle$ and  $\langle h^- \rangle$ denote
here the
mean multiplicities of $J/\psi$ mesons and negatively
charged hadrons (more than 90\% are $\pi^-$ mesons), respectively.

In the standard picture of the $J/\psi$ production
based on  the {\it hard creation } of $c\overline{c}$ pairs  
and the 
subsequent 
{\it  suppression } of the bound $c\overline{c}$  states 
the observed scaling behavior of the  $J/\psi$
multiplicity  
appears to be  due to 
an `accidental'  cancelation of  several large effects.
A very different picture was proposed in Ref.~\cite{GG} 
which explains a scaling 
property of the $J/\psi$ multiplicity 
\begin{equation}\label{scaling}
\frac {\langle J/\psi \rangle}  {\langle h^- \rangle}~\cong~ 
const(A)
\end{equation}
by assuming that a dominant fraction of 
$J/\psi$ mesons is produced directly at hadronization
according to the available hadronic phase space.
$J/\psi$ mesons are neutral and unflavored, i.e., all charges 
conserved in the strong interaction  
(electric charge, baryon number, strangeness and charm)
are equal to zero for this particle.
Therefore, its production is not influenced by the conservation laws of
quantum numbers.
Consequently, the $J/\psi$ production can be calculated
in the grand canonical approximation and, therefore, its multiplicity is
proportional
to the volume, $V$, of the matter at hadronization. 
Thus, the statistical yield 
of $J/\psi$ mesons 
at hadronization 
is  given by
\begin{eqnarray}\label{stat}
& &\langle J/\psi \rangle~=~\frac{(2j+1)~V}{2\pi^2} 
\int_0^{\infty}p^2dp~\frac{1}{\exp[(p^2+m_{\psi}^2)^{1/2}/T_H]~-~1}~\\
&\cong& \frac{(2j+1)~V}{2\pi^2} T_H  m_{\psi}^2
K_2\left(\frac{m_{\psi}}{T_H}\right) \cong 
(2j+1) V  
\left(\frac{m_{\psi}T_H}{2\pi}\right)^{3/2} \exp\left(-~ 
\frac{m_{\psi}}{T_H}\right)~, \nonumber
\end{eqnarray}
where $j=1$ and $m_{\psi} \cong 3097$~MeV are the spin and the mass 
of the $J/\psi$ meson and
$T_H$ is the hadronization 
temperature.  
The previously mentioned  results of the analysis
of hadron yield systematics  in elementary and nuclear collisions 
within the statistical approach
indicate that the hadronization temperature $T_H$ 
is the same for different
colliding systems and collision energies. This reflects
the universal
feature of the hadronization process.

The total entropy of the produced matter is proportional to its 
volume. As most of the entropy in the final state
is carried by pions, the pion multiplicity is also expected to be 
proportional to the volume of the hadronizing matter.
Thus the scaling property (\ref{scaling}) follows directly 
from the hypothesis of  statistical production of $J/\psi$  mesons
at hadronization and the universality of the parameter $T_H$.
Since elements of hadronizing matter move in the overall center
of mass system
the volume $V$ in Eq.~(\ref{stat})
characterizes in fact the sum of the proper volumes of all
elements in the collision event. 

The hypothesis of  statistical production of $J/\psi$ mesons
at a constant hadronization temperature $T_H$ leads  
to the prediction of a 
second scaling property of the $J/\psi$ multiplicity, namely:
\begin{equation}\label{scaling2}
\frac {\langle J/\psi \rangle}  {\langle h^- \rangle}~\cong~ 
const(\sqrt{s})
\end{equation}
which should be valid for sufficiently large c.m.
energies, $\sqrt{s}$.
This scaling property is illustrated  in Fig.~4
which shows the ratio $\langle J/\psi \rangle/\langle h^- \rangle$
as a function of $\sqrt{s}$ for proton--nucleon interactions.
The experimental data on $J/\psi$ yields are taken from a compilation 
given in \cite{Sch:94}. The values of $\langle h^- \rangle$
are calculated using a parameterization of the experimental 
results as proposed in \cite{Ga:91}.

Onwards from the CERN SPS energies,
$\sqrt{s} \cong 20$~GeV, the ratio 
$\langle J/\psi \rangle/\langle h^- \rangle$ 
is approximately constant, in line with the expected scaling
behavior
(\ref{scaling2}).
The rapid increase of the ratio with collision energy
observed below $\sqrt{s} \cong 20$~GeV should be attributed
to a significantly larger energy threshold for the $J/\psi$
production than for the  pion production.
In terms of the statistical approach
the effect of strict energy--momentum conservation has to be
taken into account by use of the microcanonical formulation
of the model.

The statistical $J/\psi$ multiplicity (\ref{stat})
depends on two parameters, $T_H$ and $V$. 
However,
a simple way to estimate of the crucial temperature
parameter in Eq.~(\ref{stat}) 
from the experimental data is possible, provided that
we find a second hadron  which 
has the properties of the $J/\psi$ meson, i.e.,
it is neutral, unflavored and stable with respect to strong decays.
The best candidate is the $\eta$ meson.
The multiplicity of $\eta$ mesons seems to obey also the scaling
properties (\ref{scaling}) and (\ref{scaling2}). 
The independence of the $\langle \eta \rangle/\langle \pi^0 \rangle$
ratio on the collision energy  was observed 
quite a long time ago \cite{Do:78}.
Recent data on $\eta$ production 
suggest that $\langle \eta \rangle/\langle \pi^0 \rangle$
ratio is also independent of the size of the colliding
objects.
In central Pb+Pb collisions at 158 A~GeV the ratio 
$\langle \eta \rangle/\langle \pi^0 \rangle=0.081\pm 0.014$ 
is measured \cite{Pe:98}.
It is consistent with the values of the ratio
reported for all 
inelastic p+p at 400 GeV \cite{Ag:91} (0.077$\pm$0.005)
and
S+S at 200~A$\cdot$GeV \cite{Al:95} (0.12$\pm$0.04).
From the measured ratios, 
$\langle J/\psi \rangle/\langle h^- \rangle$ and
$\langle \eta \rangle/\langle \pi^0 \rangle$, 
we estimate a mean
ratio $\langle J/\psi \rangle/\langle \eta \rangle =
(1.3 \pm 0.3) \cdot 10^{-5} $.
Here we use the experimental ratio 
$\langle \pi^0 \rangle/\langle h^- \rangle ~\cong~ 1 $ in
N+N interactions \cite{Ha:91}. 
Under the hypothesis of the statistical production of
$J/\psi$ and $\eta$ mesons at hadronization
the measured ratio can be compared to the ratio
calculated using Eq.~(\ref{stat}):
\begin{equation}\label{eta}
\frac{\langle J/\psi \rangle}{\langle \eta \rangle}~\cong~
 \frac{3~ m^2_{\psi}~ K_2(m_{\psi}/T_H)}
{m^2_{\eta}~ K_2(m_{\eta}/T_H)}~,
\end{equation}
where $m_{\eta} \cong 547$~MeV is the mass of the $\eta$ meson.
This leads to an estimate of the hadronization temperature,
$T_H\approx 176 $ MeV.
A graphical solution of Eq. (\ref{eta}) is shown in Fig.~5
which illustrates the high sensitivity of the estimate
of the $T_H$ parameter by using  the 
$\langle J/\psi \rangle/\langle \eta \rangle$ ratio.
This is due to the large difference between mass of the $J/\psi$
and the $\eta$ mesons as the right hand side of Eq.~(\ref{stat})
is approximately proportional to $\exp[(m_{\eta}-m_{\psi})/T_H]$.

In the proposed model
the creation of the $J/\psi$ mesons is due to
the  straight thermal production at hadronization and not due to
the coalescence of $c \overline{c}$ quarks produced before
hadronization.
Therefore the
yield of $J/\psi$ mesons is independent
of the production of open charm, which
is carried
mainly by the $D$ mesons
in the final state.
The $D$ mesons multiplicity is determined
by the number of $c\overline{c}$ quark pairs created in the
early {\it parton stage} before the hadronization.
Assuming chemical equilibrium of charm in the QGP stage
of A+A collision we can estimate the ratio of the open charm hadrons
to pions. 
The charm to entropy
ratio is defined by the quantity
\begin{equation}\label{rc}
R_c~\equiv~\frac{N_c+N_{\bar{c}}}{S}~,
\end{equation}
similar to Eq.~(\ref{sentr}) for the ratio of strangeness to entropy.
The behaviour of $R_c$ (\ref{rc}) for the HG and QGP
is shown in Fig.~6 as a function of $T$ for $\mu_B=0$.
In Fig.~6 the quantity $R_s$ (\ref{sentr}) from Fig.~1
is also presented for a comparison. 
The behaviour of $R_c$ is completely different from
that of $R_s$.
$R_c$ in the QGP is strongly increasing with $T$
and its values are much larger than in the HG.
The experimental analog of $R_c$ is 
$E_c=\langle D \rangle /
\langle \pi \rangle ,$
which measures the ratio of the mean multiplicities
of $D$ mesons to the multiplicity of
pions.
The assumption of the conserved
total entropy and
the total number of charm quarks and antiquarks 
during the hadronization of QGP
leads then to the picture of chemical non-equilibrium
hadron gas at the chemical freeze-out with a strong enhancement
of $D$ mesons yield. The ratio of $D$ mesons to pions should 
strongly increase with the collision energy from the SPS to 
the RHIC.

\vspace{0.5cm}
\noindent
{\bf IV. Summary}

Statistical production of starngeness and charm 
discussed in this talk can be summarized as the following:
\begin{itemize}
\item 
The transition to the QGP in A+A collisions
could be seen as a non-monotonic dependence 
on collision energy of the strangeness to pion ratio in the energy region
between the AGS and the SPS.
\item
The yield of $J/\psi$ mesons at the CERN SPS 
can be understood assuming the statistical   
production of $J/\psi$ at the hadronization,
and it is sensitive to the hadronization temperature.
The $D$ mesons multiplicity is determined by the number of
$c\overline{c}$ quark pairs created in the early QGP stage.
The ratio of $D$ mesons to pions is expected to
increase strongly with the collision energy from the SPS to the RHIC.
\end{itemize}

\vspace{0.5cm}
\noindent
{\bf Acknowledgements}\\
I thank Marek Ga\'zdzicki and Grandon Yen for
the fruitful collaboration.
 This work is supported by  DFG, Germany.

\begin{figure}[t]\label{fig1}
\mbox{}
\begin{center}
\vfill
\leavevmode
\epsfysize=15cm \epsfbox[100 160 540 760]{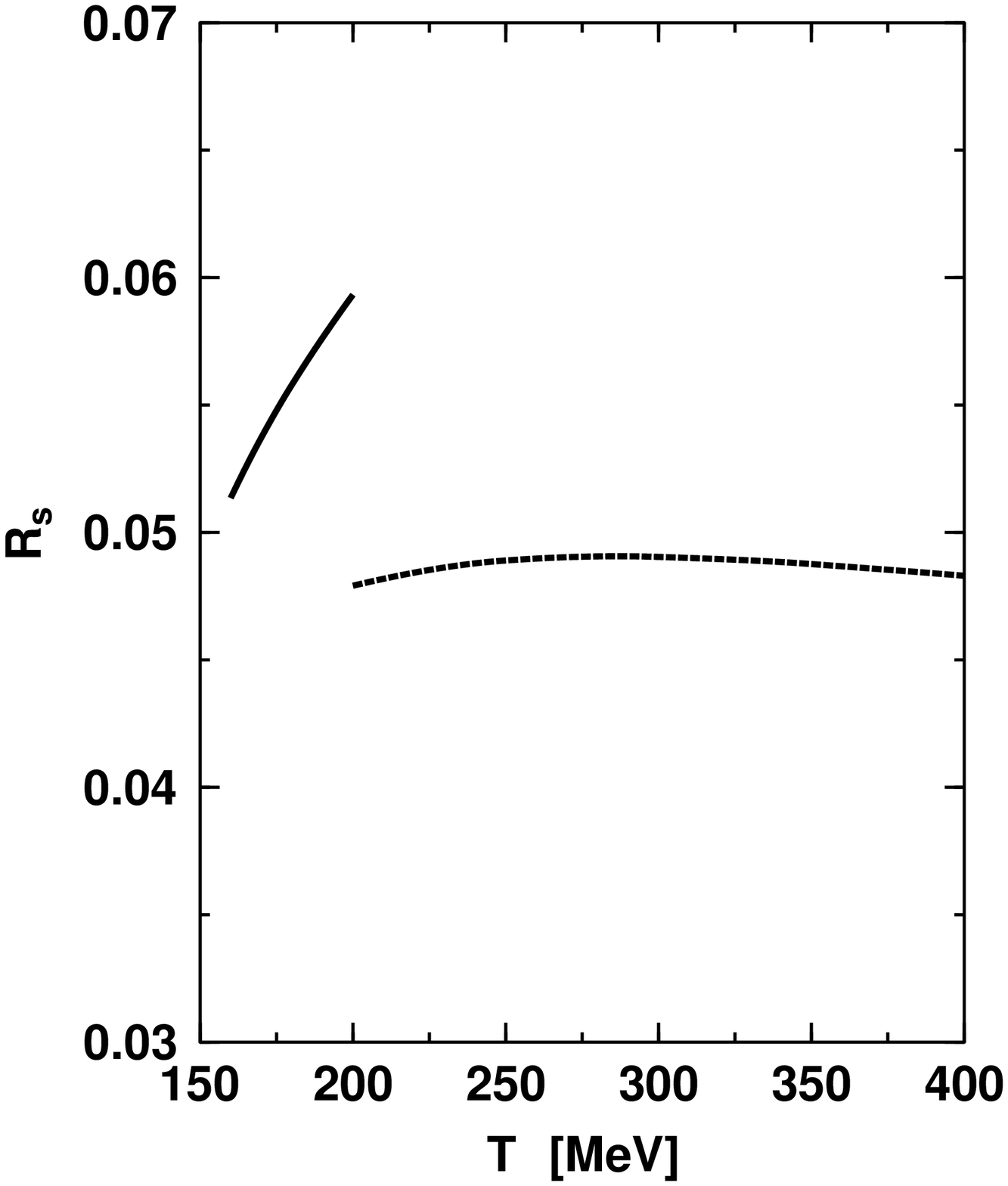}
\vfill
\caption{
$R_s$ (1) at $\mu_B=0$ for the HG (solid line) and the QGP (dashed line).
}
\end{center}
\end{figure}

\begin{figure}[t]\label{fig2}
\mbox{}
\begin{center}  
\vfill
\leavevmode
\epsfysize=14cm \epsfbox[100 1 500 500]{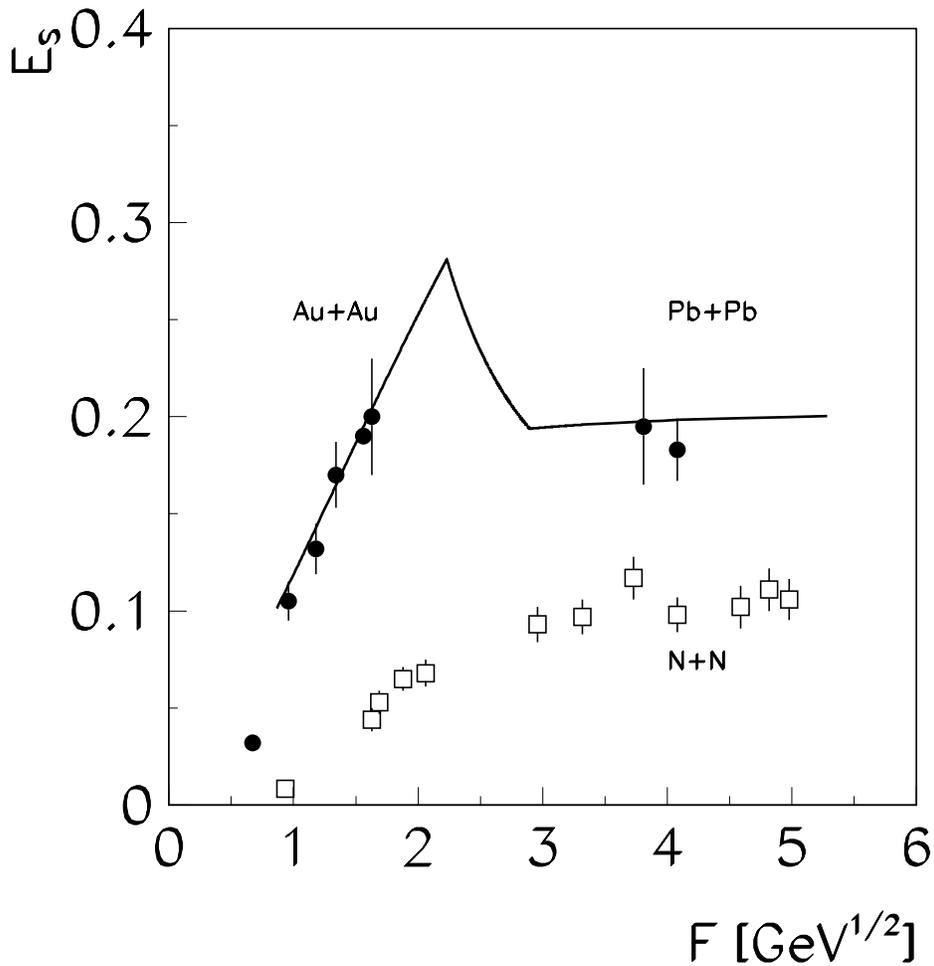}
\vfill
\caption{
The dependence of the strangeness to pion ratio, $E_s$ (2), for the
central A+A collisions (closed circles) and nucleon-nucleon
interactions (open squares) as a function of the collision energy
measured by the variable $F= (\sqrt{s} - 2m_N)^{3/4}/\sqrt{s}^{1/4}$.
The solid line shows predictions of the statistical model
of Ref.~\cite{Ga1:98}. A transition to the QGP takes place between
the AGS ($F\approx 2$) and the SPS ($F\approx 4$) energies.
}
\end{center}
\end{figure}

\begin{figure}[t]\label{fig3}
\mbox{}
\begin{center}
\vfill
\leavevmode
\epsfysize=14cm \epsfbox[100 1 500 500]{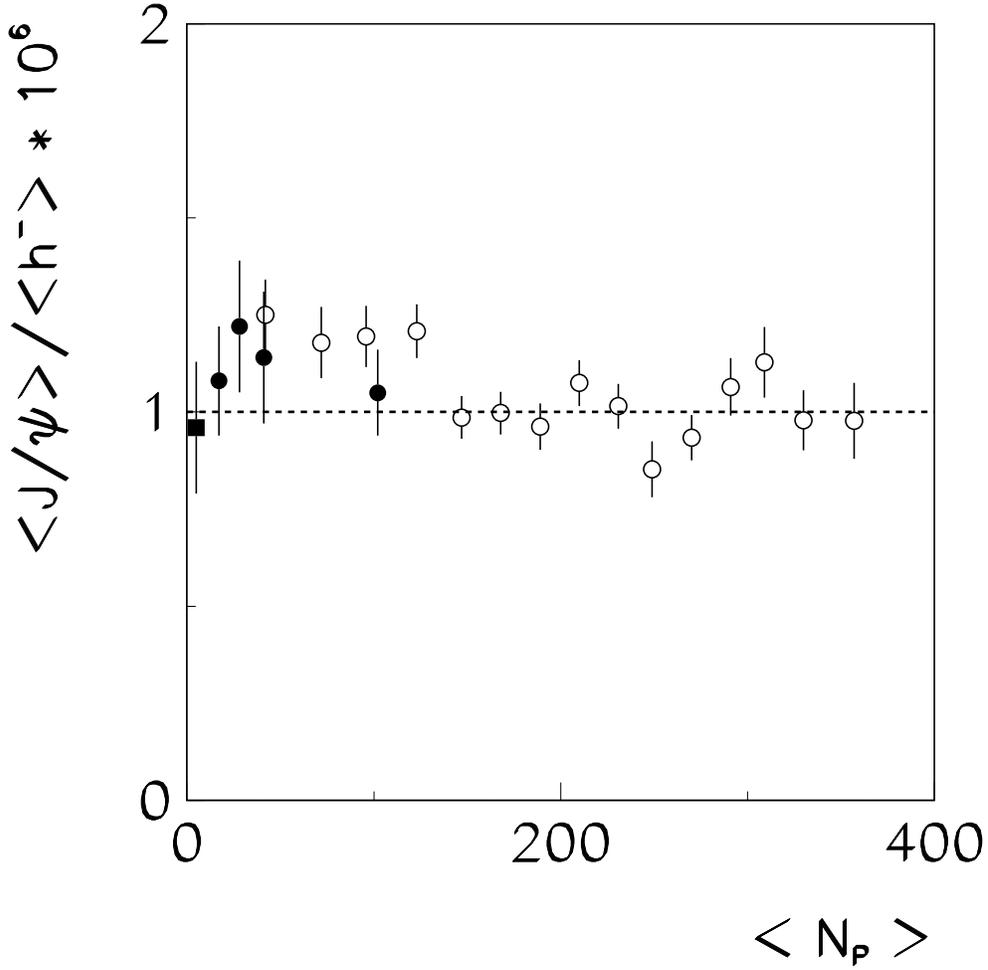}
\vfill   
\caption{
The ratio of the mean multiplicities of $J/\psi$ mesons
and negatively charged hadrons for inelastic nucleon--nucleon (square) and
inelastic O+Cu, O+U, S+U and Pb+Pb (circles) interactions at
158 A$\cdot$GeV plotted as a function of the mean
number of participant nucleons.
The results for Pb+Pb interactions measured for different
centralities of collisions are shown by open circles.
For clarity the N+N point is shifted from
$\langle N_P \rangle = 2$ to $\langle N_P \rangle = 5$.
The dashed line indicates the mean value of the ratio. 
}
\end{center}
\end{figure}

\begin{figure}[t]\label{fig4}
\mbox{}
\begin{center}
\vfill
\leavevmode
\epsfysize=15cm \epsfbox[100 1 500 500 ]{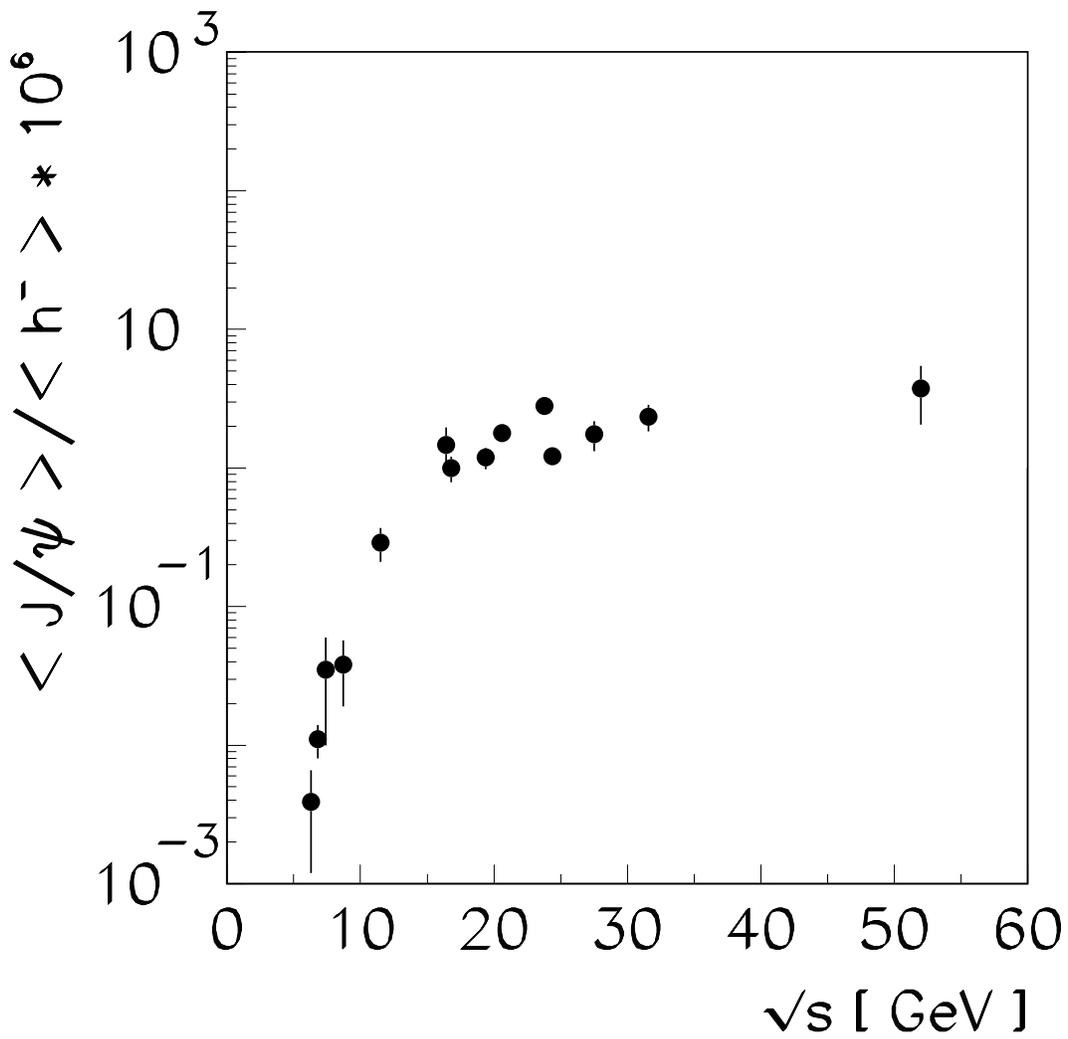}
\vfill
\caption{
The ratio of the mean multiplicities of $J/\psi$ mesons
and negatively charged hadrons for inelastic proton--nucleon
interactions  as a function of the collision energy in the
center of mass system.
}
\end{center}
\end{figure}

\begin{figure}[t]\label{fig5}
\mbox{}
\begin{center}
\vfill   
\leavevmode
\epsfysize=15cm \epsfbox[100 1 500 500]{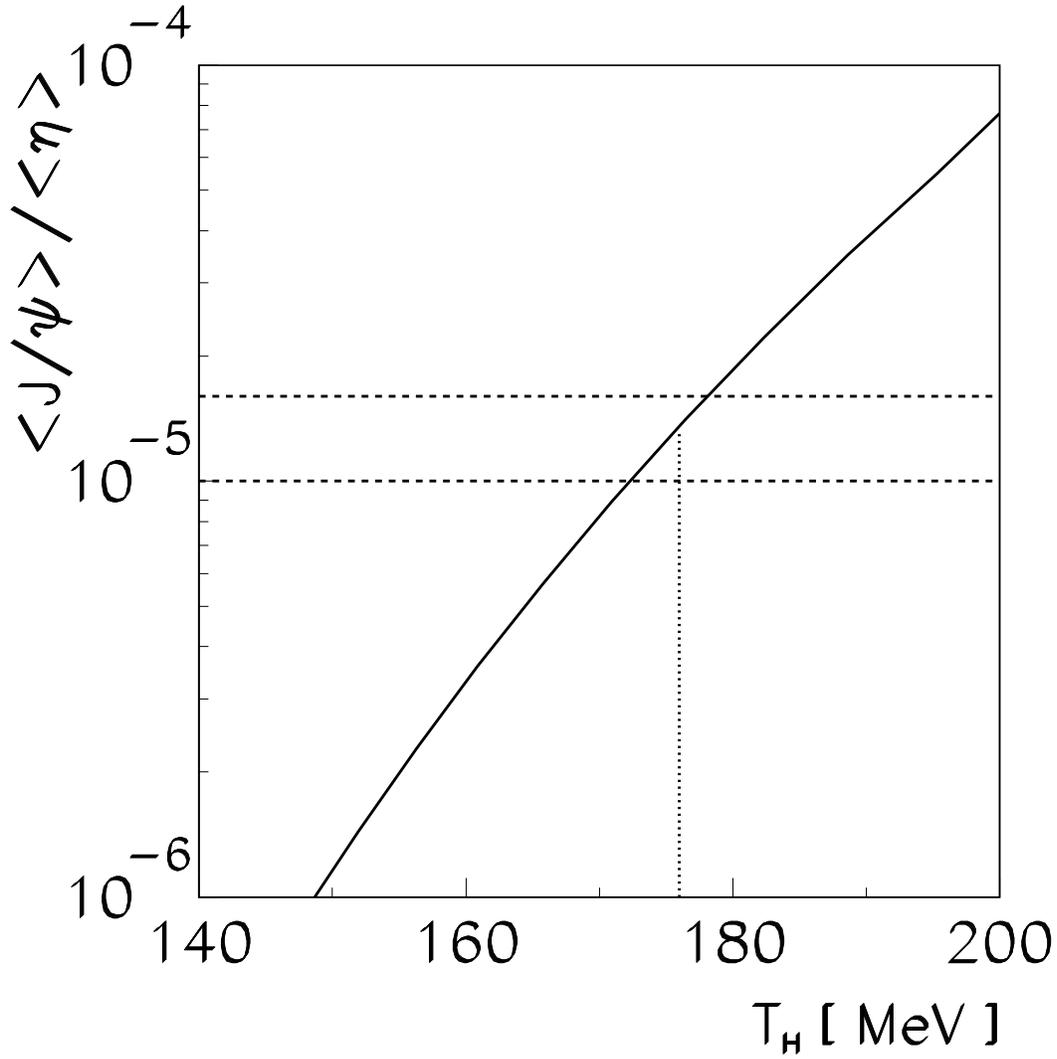}
\vfill
\caption{
The $\langle J/\psi \rangle/\langle \eta \rangle$ ratio
calculated under hypothesis of the statistical production of
$J/\psi$ and $\eta$ mesons at hadronization (solid line)
as a function of the hadronization temperature.
Band shown by dashed lines is drawn at $\pm \sigma$
around the mean experimental value of the
$\langle J/\psi \rangle/\langle \eta \rangle$ ratio.
The dotted line indicates $T_H= 176$ MeV.
}
\end{center}
\end{figure}

\begin{figure}[t]\label{fig6}
\mbox{}
\begin{center}
\vfill
\leavevmode
\epsfysize=15cm \epsfbox[100 160 540 760]{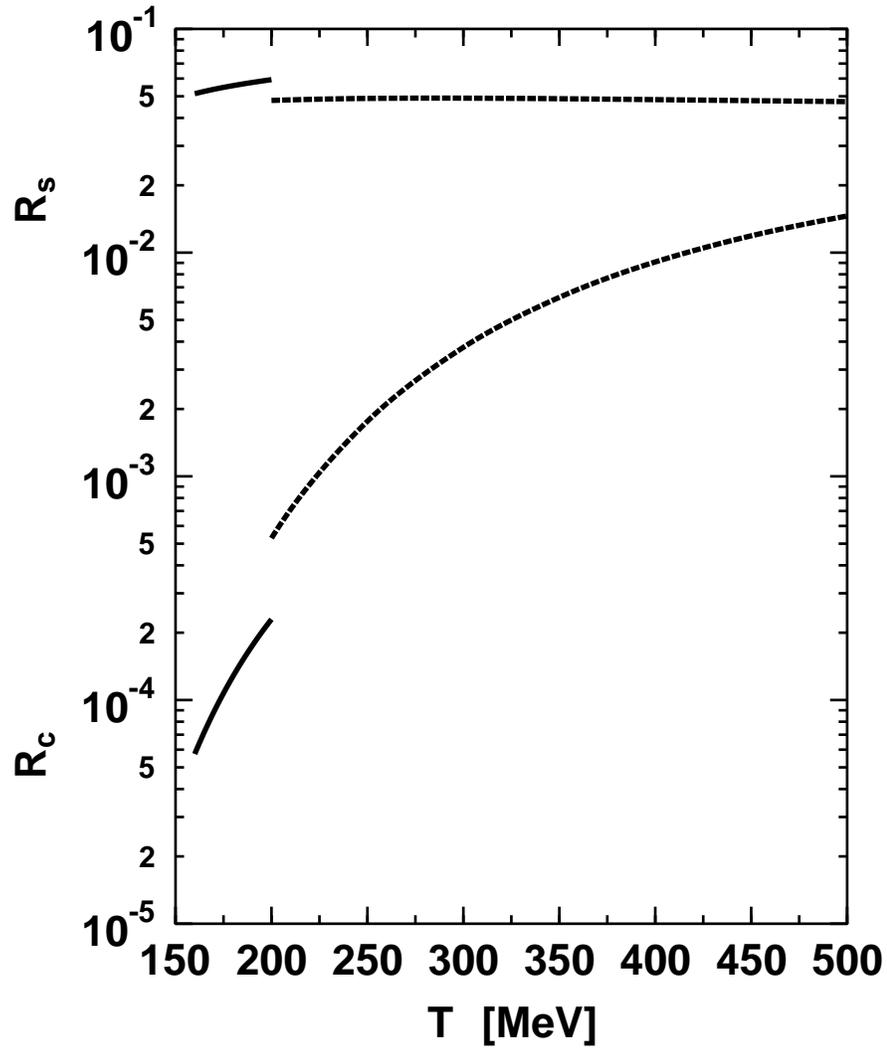}
\vfill   
\caption{  
$R_c$ (\ref{rc}) at $\mu_B=0$ for the HG (solid line) and the QGP (dashed
line). $R_s$ behaviour from Fig.~1 is also presented
(upper lines) for a comparison.
}
\end{center}
\end{figure}

\end{document}